\documentclass[aps,pra,amsmath,amssymb,
 reprint,superscriptaddress]{revtex4-1}
\usepackage{graphicx}
\usepackage{dcolumn}
\usepackage{bm}

\draft 

\begin{document}





\title{Sensitivity limits of a Raman atom interferometer as a gravity gradiometer} 








\author{F. Sorrentino}
\author{Q. Bodart}
\affiliation{Dipartimento di Fisica e Astronomia \& LENS, Universit\`a di Firenze, INFN Sezione di Firenze, via Sansone 1, I-50019 Sesto Fiorentino (FI), Italy}
\author{L. Cacciapuoti}
\affiliation{European Space Agency, Research and Scientific Support Department, Keplerlaan 1, 2200 AG Noordwijk, The Netherlands}
\author{Y.-H. Lien} 
\affiliation{Centre for Cold Matter, Department of Physics, Imperial College of
London, London, SW7 2BW, UK}
\author{M. Prevedelli}
\affiliation{Dipartimento di Fisica dell'Universit\`a di Bologna, Via Irnerio 46, I-40126, Bologna,  Italy}
\author{G. Rosi}
\author{L. Salvi}
\author{G. M. Tino}\email[]{guglielmo.tino@fi.infn.it}
\affiliation{Dipartimento di Fisica e Astronomia \& LENS, Universit\`a di Firenze, INFN Sezione di Firenze, via Sansone 1, I-50019 Sesto Fiorentino (FI), Italy}




%





\date{\today}

\begin{abstract}

We evaluate the sensitivity of a dual cloud atom interferometer to the measurement of vertical gravity gradient. We study the influence of most relevant experimental parameters on noise and long-term drifts. Results are also applied to  the case of doubly differential  measurements of the gravitational signal from local source masses. We achieve a short term sensitivity of $3\times 10^{-9}$\,g/$\sqrt{\hbox{Hz}}$ to differential gravity acceleration, limited by the quantum projection noise of the instrument. Active control of the most critical parameters allows to reach a resolution of $5\times 10^{-11}$\,g after 8000\,s  on the measurement of differential gravity acceleration. The long term stability is compatible with a measurement of the gravitational constant $G$ at the level of $10^{-4}$ after an integration time of about 100 hours.


\end{abstract}


\maketitle 



\section{Introduction}
\label{intro}



Atom interferometry provides extremely sensitive and accurate tools for the measurement of inertial forces, finding important applications both in fundamental
physics and applied research \cite{Cronin09,Tino2013}. 
Quantum sensors based on atom interferometry 
had a rapid development during the last two decades, and are expected to play a crucial role for
science and technology in the next future.

The performances of atom
interferometry sensors have been already demonstrated in the measurements of gravity
acceleration \cite{Kasevich92,Peters99,Mueller08,LeGouet08}, Earth's gravity gradient \cite{Snadden98,McGuirk02,Bertoldi06}, and rotations
\cite{Gustavson97,Gustavson00,Canuel06,Gauguet09}. 
Experiments based on atom interferometry
are currently running to test the Einstein's Equivalence Principle \cite{Fray04,Dimopoulos08GR}, to measure the Newtonian gravitational constant G \cite{Bertoldi06,Fixler07,Lamporesi08} and the fine structure constant $\alpha$ \cite{Bouchendira11,Lan13}, and to test fundamental physics effects in atomic systems \cite{Jacquey07,Gillot13}, while 
experiments testing general relativity \cite{Mueller08,Dimopoulos08GR,Mueller10} and the $1/r^2$ Newton 's law 
\cite{Tino02,Wolf07,Ferrari06,Sorrentino09} or searching for quantum gravity effects \cite{Camelia09} and for gravitational waves detection
\cite{Tino04,Dimopoulos09,Tino2011} have been proposed.
Accelerometers based on atom interferometry have
been developed for several applications including
metrology, geodesy, geophysics, engineering prospecting and
inertial navigation \cite{McGuirk02,Peters01,Leone06,deAngelis09,Geneves05}. 

While the sensitivity of such quantum inertial sensors has not yet reached its ultimate limits, recent progresses in atom optics are expected to yield further improvements by some orders of magnitude by increasing the momentum transfer during the interferometer sequence \cite{Mueller09,Chiow11}. These instruments are expected to reach their ultimate sensitivity in space where free fall conditions allow  very long interrogation times. \cite{Tino07,Turyshev07,Geiger11,Muntinga13,TinoQWEP2013}.

One of the most interesting features of atom interferometry sensors, besides their   sensitivity, is the ability to control systematic effects. This in turn follows from the possibility to use the quantum nature of atom-light interactions as a tool to control several sources of biases.
This makes atom interferometry sensors particularly suited for applications requiring long term stability and accuracy.

In this paper we analyze the influence of the most relevant experimental parameters on the stability and accuracy of our apparatus for gravity gradient measurements by atom interferometry. We also consider a specific experimental configuration for the measurement of the gravitational signal from local source masses. In this way, we show that the present state of our experiment is compatible with the measurement of the gravitational constant $G$ with a precision of $10^{-4}$. 

The paper is organized in the following way: section \ref{apparatus} describes the experimental apparatus and the measurement scheme of our gradiometer; section \ref{sensitivitylimits} describes how to extract differential acceleration measurements from the instrument raw data; in section \ref{parameters} we analyze the effect of most relevant experimental parameters on the gravity gradient and $G$ measurements; finally, section \ref{measurements} describes the sensitivity and long term stability performance of our apparatus.

\section{Experimental apparatus}
\label{apparatus}

In the following we present the measurement principle of our apparatus. An extensive description is given in \cite{Fattori03,Lamporesi08,Sorrentino10}. 

Our experiment is based on a dual atom interferometer, to measure the differential acceleration between two clouds of cold rubidium atoms in free fall, and on a well characterized set of source masses, to produce a controlled gravity acceleration at the location of the atomic probes. 


The atom gravimeter is based on Raman light-pulse interferometry \cite{Kasevich92}: atoms are first launched vertically
in a fountain configuration, and then illuminated by a sequence of 
light pulses acting as beam splitters and mirrors for the atomic
wave packets. The light pulses are generated by two vertically aligned and counter-propagating
laser beams,
inducing two-photon
Raman transitions between 
the hyperfine levels of the Rb ground state. 
An atom optics beam splitter consists in a $\pi/2$-pulse with length $\tau=\pi/2\Omega$, where $\Omega$ is the two-photon Rabi frequency, which drives the atom wavefunction into an equal superposition of the two hyperfine states. An atom optic mirror consists in  a $\pi$-pulse with length $\tau=\pi/\Omega$, swapping the atomic populations between the two hyperfine states. Since the two laser beams are counter-propagating, the Raman transitions result in a momentum exchange by an amount of $\hbar k_{e} = \hbar(k_1 + k_2)$, where $k_1$ and $k_2$ are the wave numbers of the two Raman laser fields. 
The atom interferometer is composed of a sequence of three Raman pulses separated by two equal time intervals $T$, i.e. a $\pi/2$-pulse to split, a $\pi$-pulse to redirect, and a $\pi/2$-pulse to recombine the atomic wavefunction. 

At the output of the interferometer, the probability
of detecting the atoms in the original hyperfine state 
is given by
$P_a=(1-\cos\phi)/2$, where $\phi$ is the phase difference
accumulated by the wave packets along the two
interferometer arms. In the presence of a uniform gravity field,
the phase shift $\phi=k_{e}gT^2$ is proportional to 
 the gravitational acceleration $g$.
 The
gravity gradiometer is obtained by  operating two simultaneous gravimeters with two vertically separated atomic clouds illuminated by the same Raman laser pulses. 
This configuration provides a measurement of the differential acceleration
between the two samples with an excellent common-mode rejection of vibration noise.

\begin{figure}
\begin{center}
\includegraphics[width=0.55\textwidth]{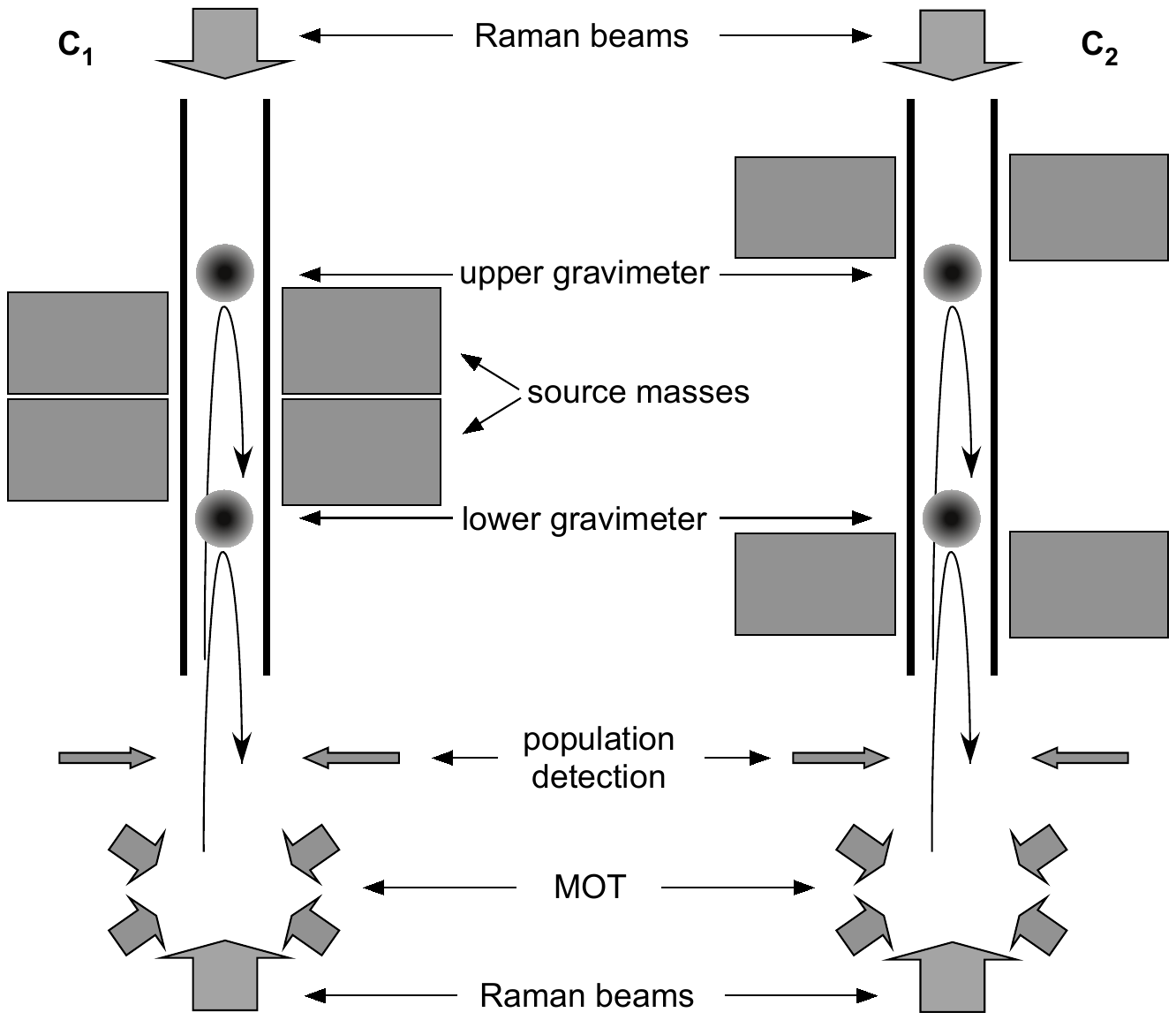}
\caption{\label{magiascheme} Scheme of the gravity gradiometer (from \cite{Sorrentino10}). $^{87}$Rb atoms are first loaded in the magneto-optical trap (MOT), and then launched vertically
in the vacuum tube with the moving optical molasses
method. Around the
apogees of the atomic trajectories, the atoms are illuminated by a sequence of laser pulses for the Raman interferometry
scheme. External source masses are typically positioned in
two different configurations ($C_1$ and $C_2$) and the induced phase shift
is measured as a function of masses positions.}
\end{center}
\end{figure}

A scheme of our gravity gradiometer is shown in figure \ref{magiascheme} with the two typical configurations of the
source masses ($C_1$ and $C_2$). We collect the $^{87}$Rb atoms in a magneto-optical trap (MOT) at the bottom of
the apparatus.
We launch the samples with the moving molasses technique along the symmetry axis of
the vacuum tube, at a temperature of about
 $2.5\,\mu$K. For the gravity gradiometer we employ two atomic
clouds  simultaneously reaching the apogees of their
ballistic trajectories at about 60\,cm and 90\,cm above the MOT.
We 
cope with the short time delay between the two launches ($\sim80$\,ms) by juggling the atoms loaded
in the MOT \cite{Legere98}. In this way we are able to launch about $10^9$ atoms in each cloud.
Shortly after launch the atoms enter the magnetically shielded vertical tube shown in figure \ref{magiascheme}, where a uniform magnetic field of 29\,$\mu$T along the vertical direction defines the quantization
axis. The field gradient along this axis is lower than
$10$\,$\mu$T/m (see section \ref{bpulse}). 
At this stage, the two atomic samples are first simultaneously addressed with a combination of a Raman $\pi$ pulses and resonant blow-away laser pulses to select a narrow velocity class and to prepare the atoms in the 
 $(F=1, m_F=0)$
state. 
The Raman lasers propagate along the vertical direction from the bottom, and are retro-reflected on a mirror above the vacuum tube. The atom interferometry sequence  takes place around the apogee of the atomic trajectories, with a sligth asymmetry to avoid double resonance at the central $\pi-$pulse (see section \ref{bpulse}).
We complete the experimental cycle by measuring the normalized population of the ground state $F=1,2$ hyperfine
levels via fluorescence spectroscopy in a chamber placed just above the MOT.

In both of the simultaneous atom
interferometers the
local acceleration is measured with respect to the common reference
frame identified by the wave fronts of the Raman lasers.
As a result, any phase noise induced by vibrations on
the retro-reflecting mirror can be efficiently rejected as common mode: 
when
plotting the signal of the upper accelerometer versus the lower one, experimental points distribute along an
ellipse. The differential phase shift $\Phi=\phi_{u}-\phi_{l}$, which is proportional to the gravity gradient, is then obtained from
the eccentricity and the rotation angle of the ellipse best fitting
the experimental data \cite{Foster02}.

For the measurement of the gravitational signal from local source masses, as for the determination of $G$, the gravity gradient measurement is repeated in the two different configurations of source masses shown in figure \ref{magiascheme}. In this way, we are able to isolate the effect of source masses from other biases introduced by Earth's gravity gradient, Coriolis forces, etc. The position of source masses is modulated between the two configurations shown in figure \ref{magiascheme} with a period $T_{mod}\simeq 15\div30$\,min.

\section{Short-term sensitivity in gravity gradient measurements}
\label{sensitivitylimits}

\begin{figure}
\includegraphics[width=0.5\textwidth]{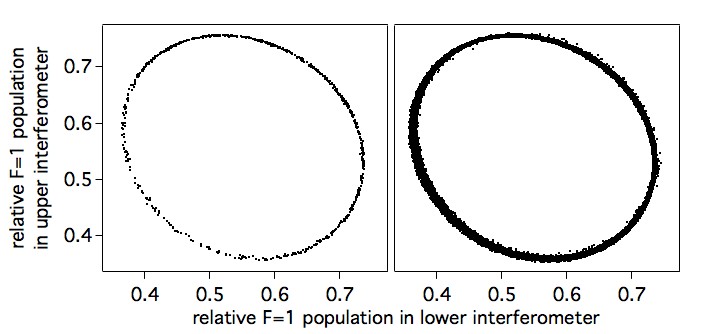}%
\caption{\label{twoellipses}Two experimental ellipses obtained with 720 (left) and 79000 (right) experimental points; each point is acquired in 1.9\,s.
}%
\end{figure}

The differential gravity acceleration is calculated from the phase angle of an ellipse whose points  $ (x, y)$ are the fraction of atoms in the $F=1$ state of each cloud as measured from the fluorescence signals.
Figure \ref{twoellipses} shows two typical elliptical plots  for 720 and $\sim20000$ experimental points. 
We use a least squares fitting algorithm to extract the differential phase $\Phi$ of the gradiometer. We fit the experimental data to the parametric equations 

\begin{equation}
\label{ellipseq}
\begin{cases}
x(t)=A\sin(t)+B\\
y(t)=C\sin(t+\Phi)+D\\
\end{cases}
\end{equation}
where the $A$ and $C$ parameters represent the amplitudes of the 
interference fringes for the upper and lower interferometers, and $(B,D)$ are the coordinates of the ellipse center
(see eq.~(\ref{eggequation}) below). Although more sophisticated algorithms have been proposed to retrieve $\Phi$ with Bayesian estimators \cite{Stockton07}, least-squares ellipse fitting is adequate for the analysis of sensitivity and long term stability \cite{Sorrentino10}.

\subsection{Detection noise and quantum projection noise}
\label{detectionnoise}

The sensitivity to gravity gradient measurement can be modeled by including noise terms in eq.~(\ref{ellipseq}), i.e. a term $\delta\Phi$  describing differential phase fluctuations, two terms $\delta A$ and $\delta C$ describing fringe contrast fluctuations, and two terms $\delta B$ and $\delta D$ describing fringe bias fluctuations; moreover, we add two terms $\delta x_d(t)$ and $\delta y_d(t)$ respectively to the lines of eq. (\ref{ellipseq}), describing additive detection noise. The total fluctuations $\delta x(t)$ and $\delta y(t)$ of the atom interferometry signals depend on the parameter $t$; by taking the average over $t$ 

\begin{equation}
\label{noiseterms}
\begin{cases}
\langle\delta x^2\rangle = \frac{1}{2}\langle\delta A^2\rangle + \langle\delta B^2\rangle + \langle\delta x_{d}^2\rangle\\
\langle\delta y^2\rangle = \frac{1}{2}\langle\delta C^2\rangle + \langle\delta D^2\rangle + \langle\delta y_{d}^2\rangle+\frac{C^2}{2}\langle\delta \Phi^2\rangle\\
\end{cases}
\end{equation}
The different contributions are not easily disentangled experimentally; in this section we give a model for detection noise, and in the following section we discuss the effect of contrast and bias fluctuations.

\begin{figure}
\includegraphics[width=0.5\textwidth]{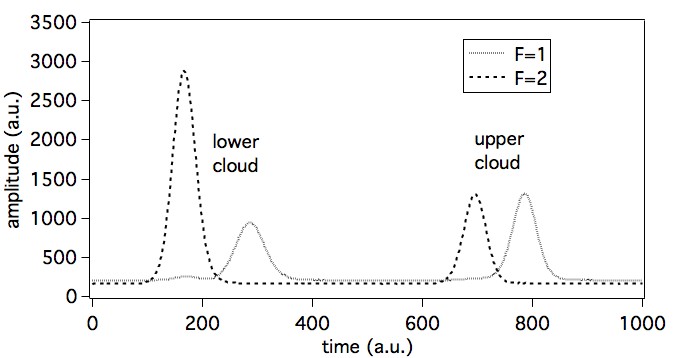}%
\caption{\label{fitpeaks} Typical plot of detection signals after the atom interferometry sequence; the two curves are for the $F=1$ and $F=2$ channels respectively; for each curve, the two peaks are for the lower and upper cloud, respectively.
}%
\end{figure}

The detection signals are obtained by collecting the atomic fluorescence from the two hyperfine states in two separate regions (see section \ref{detection}) using independent photodiodes. Typical photodiode signals are shown in figure \ref{fitpeaks}. The population $n_{ij}$ of the $F = i$ state ($i=1,2$) is proportional to the area $A_{ij}$ of the corresponding peak in the detection signal, i.e. $n_{ij}=\eta_{i}A_{ij}$, where $j=1,2$ labels the upper and lower cloud, respectively.

In general, the detection noise is not uniform along the ellipse, because the populations $n_{ij}$ depend on $t$. Let us assume for simplicity that the detection efficiency is the same for the two channels, i.e. $\eta_1=\eta_2$. Since the signal $x(t)$ in eq.~(\ref{ellipseq}) is given by the normalized population $n_{11}/(n_{11}+n_{21})$, the detection noise can be written as

\begin{equation}
\label{detectionnoiseq}
\delta x_{d}^2(t) = \frac{x^2(t)\delta n_{21}^2+[1-x^2(t)]\delta n_{11}^2}{n_x^2}
\end{equation}
where $n_x=n_{11}+n_{21}$, and $\delta n_{11}$  ($\delta n_{21}$) is the detection noise for $F=1$ ($F=2$) atoms. A fundamental lower limit to $\delta x_{d}$ is given by  the quantum projection noise (QPN) $\delta n_{ij}^2=n_{ij}$. In this case eq.~(\ref{detectionnoiseq}) reads
$\delta x_{d}^2(t)=x(t)[1-x(t)]/n_x$; applying eq.~(\ref{ellipseq}) and averaging over $t$ we obtain

\begin{equation}
\label{qpnequation}
\begin{cases}
\langle\delta x_{QPN}\rangle=\sqrt{\frac{2B(1-B)-A^2}{2n_x} }\\
\langle\delta y_{QPN}\rangle=\sqrt{\frac{2D(1-D)-C^2}{2n_y}}\\
\end{cases}
\end{equation}
In our typical experimental conditions, $n_x\simeq n_y\simeq2\times10^5$\,atoms, 
$A\simeq C\simeq0.225$, $B\simeq D\simeq0.5$, thus
the noise per shot amounts to $\delta x_{\rm{QPN}}\simeq\delta y_{\rm{QPN}}\simeq0.0011$. We investigated the QPN limit to the gravity gradient measurements with a numerical simulation: we generated several  ellipses described by eq.~(\ref{ellipseq}), where $t$ is uniformly distributed in $[0;\pi]$ and adding 
Gaussian noise to each line, with standard deviation 0.0011. We calculated the Allan variance for $\Phi$ as resulting from least-square fitting of simulated ellipses with contrast and bias close to our typical values. The Allan variance $\sigma_\Phi(N)$ drops as the square root of the number $N$ of points, $\sigma_\Phi(N)=0.015/\sqrt{N}$; we repeated the simulation for different values of  $A\simeq C$, and verified that $\sigma_\Phi(N)$ scales with the inverse of the contrast.

Detection noise can be larger than the QPN limit due to technical noise sources such as intensity and frequency fluctuations of probe laser beams, electronic noise, stray light etc.
In our setup, an upper limit to technical detection noise can be estimated from the fit  of the detection peaks (see figure \ref{fitpeaks}). After removing a small amount of crosstalk between channels, we fit each  peak to the product of a Gaussian and a fourth order polynomial

\begin{equation}
\label{detectionfunction}
h(1+\alpha x+\beta x^2 + \gamma x^3 + \delta x^4)\exp{\left[-\frac{(x-x_0)^2}{2\sigma^2}\right]}+ B
\end{equation}
in order to account for the signal distortion due to the finite bandwidth of the photodiode, so that the area $A_{ij}$ of a peak is given by 

\begin{equation}
\label{area}
A_{ij}=h\sigma\sqrt{2\pi}(1+\beta \sigma^2 + 3 \delta \sigma^4)
\end{equation}
%

We collect the atomic fluorescence with two large area photodiodes (Hamamatsu S7510, active area 11$\times$6 mm$^2$), with 1 G$\Omega$ transimpedance amplifiers. The bandwidth is of the order of a few kHz. 
We optimize the noise and bandwidth by bootstrapping the large capacitance of the photodiode with a low noise JFET, as described in \cite{boottrapping}. In this way we reach a current noise level of $\sim7$\,fA/$\sqrt{\hbox{Hz}}$ limited by the Johnson noise of the 1\,G$\Omega$ resistor and the photodiode dark current. In our typical conditions, RMS fluctuations on peak height and width are $\delta h/h\sim\delta\sigma/\sigma\sim0.004$ and the noise $\delta A_{ij}/A_{ij}\sim\delta n_{ij}/n_{ij}\sim0.006$ corresponds to the QPN limit for $\sim30000$ atoms.  Thus technical noise is smaller than QPN by about a factor 2.

\subsection{Noise on ellipse contrast and bias}
\label{contrastbias}

Sensitivity and long term stability of the gravity gradient measurement can be limited by noise in the $x$ and $y$ signals, by fluctuations and/or drifts in the contrast and center of the ellipses, and by sources of instability of the $\Phi$ value itself. The main sources of instability in ellipse contrast, bias and phase angle are discussed in  section \ref{parameters}. Let us call $t_{e}$ the measurement time to acquire an ellipse. As shown in \cite{Sorrentino10}, it is possible to obtain a reliable value for $\Phi$ with an ellipse containing a few hundreds of points. We typically use $100\div700$ points per ellipse, corresponding to a measurement time $t_{e}\sim190\div1200$\,s. In our typical experimental conditions, fluctuations $\delta\Phi$ of the differential phase on time scales shorter than $t_e$ are negligible (see section \ref{long}). On the other hand, the slow changes in the $A$, $B$, $C$ and $D$ parameters occurring on a time scale longer than $t_{e}$, as visible on the right of figure \ref{twoellipses} , are efficiently rejected. The short term sensitivity will be mainly determined by detection noise, and possibly by fast fluctuations of ellipse contrast and position, such as those caused by changes in the detection efficiency (see section \ref{detection}) or in the Raman laser power (see section \ref{Raman}). Contrast and bias fluctuations on times longer than $t_{e}$ do not affect the long term stability of gravity gradient measurement, which is thus only limited by slow distortions and rotations of the ellipse, such as those from Coriolis acceleration (see section \ref{Coriolis}) or detection efficiency changes (see section \ref{detection_efficiency}). The following section provides a systematic characterization of the influence of most relevant experimental parameters on ellipse contrast, bias and rotation angle.

\section{Long-term stability and accuracy: impact of most relevant experimental parameters}
\label{parameters}

Noise sources which equally affect the upper and lower atom interferometer (i.e. vibrations, tidal effects, relative phase noise of Raman lasers, etc.) are rejected as common mode in the gravity gradient measurement. In the following subsections, we will investigate those experimental parameters which affect the two atom interferometers differently; such parameters can in principle limit the sensitivity and long term stability of gravity gradient measurements. 
Due to the double differential scheme, the measurement of gravity signal from local source masses is even more robust with respect to noise cancellation and control of systematic effects. Indeed, the only effects which can affect the  
   measurement of local source masses are those which either depend on the position of source masses, or change on a time scale shorter than the cycling time $T_{mod}$ of masses positions. 

We separately investigated the effect of various parameters. We recorded the ellipse phase angle in the two configurations of source masses, $\Phi_{C1}$ and $\Phi_{C2}$, for different values of each parameter $\alpha$; for each value of the parameter, we calculated the average ellipse angle $\bar\Phi=(\Phi_{C1}+\Phi_{C2})/2$ and the difference $\Delta\Phi=\Phi_{C1}-\Phi_{C2}$.
 From $\bar\Phi(\alpha)$ we can deduce requirements on the stability of the parameter $\alpha$ on time scales shorter than $T_{mod}$ for 
  the measurement of local source masses, as well as on the long term stability of $\alpha$ for gradient measurements; from $\Delta\Phi$ we can deduce requirements on the long term stability of $\alpha$ for  
  the measurement of local source masses.

\subsection{Intensity fluctuations of cooling laser}
\label{cooling}

\begin{figure}
\includegraphics[width=0.5\textwidth]{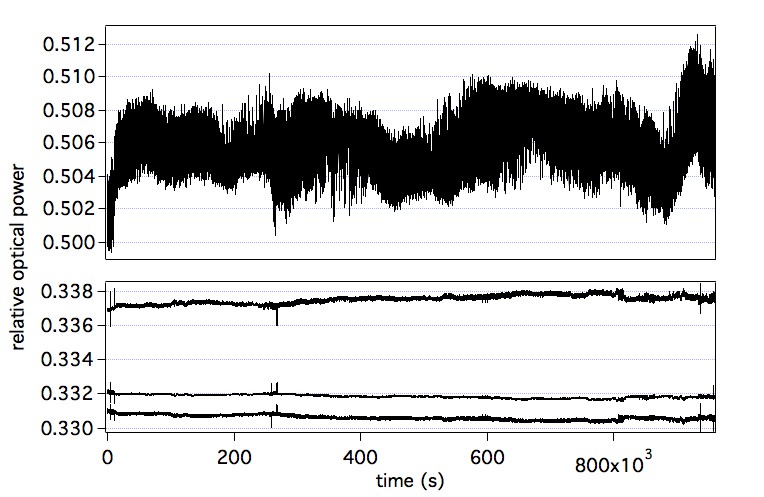}
\caption{\label{coolingfluctuations} Time fluctuations of MOT laser intensities; the upper plot shows the relative reading of the two photodiodes at the input of the $1\to 3$ splitters (see text) over about eleven days; the lower plot shows the relative readings of the three photodiodes monitoring the output of MOT collimators within the triplet generated from the ``down'' beam (see text) over the same time interval.}%
\end{figure}

The total power and intensity ratio of the six MOT laser beams affect the number of atoms as well as the temperature and launching direction in the atomic fountain. Such effects may influence the upper and lower interferometers differently. With the chosen launch configuration in the atomic fountain, the six MOT beams are produced in two independent triplets from the output of a single MOPA (Master Oscillator Power Amplifier). The MOPA output 
is split in two parts, the ``up'' and ``down'' beams, which are separately controlled in frequency and amplitude with two AOMs. Each beam is then coupled into a polarization maintaining (PM) optical fiber, and sent to a $1\to 3$ fiber splitter to produce a triplet of beams, which are delivered through PM fibers to collimators attached to the MOT chamber. In this configuration, intensity fluctuations of MOT laser beams are dominated by changes in the AOM and fiber coupling efficiency of the ``up'' and ``down'' beams in the optical bench. Fluctuations generated in the $1\to 3$ splitters are negligible. This is shown in figure \ref{coolingfluctuations}, where the typical power fluctuations of the ``up'' and ``down'' beams at the input of the $1\to 3$ splitters are compared with the relative intensity fluctuations, at the output of the MOT collimators, within the triplet generated from the ``down'' beam. The RMS fluctuations of intensity ratio between ``up'' and ``down'' beams is larger than the relative intensity fluctuations within each triplet by one order of magnitude.

\begin{figure}
\includegraphics[width=0.5\textwidth]{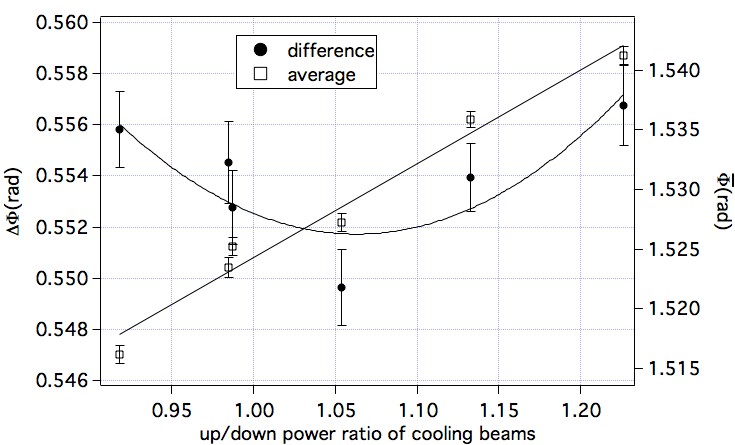}%
\caption{\label{cool}Average and differential ellipse angle for the two configuration of source masses, versus the power ratio between upper and lower cooling laser beams. Solid lines are least squares parabolic (black points) and linear (white squares) fits to the data.}%
\end{figure}

For this reason, we can restrict the analysis to the effect of intensity fluctuations at the input of the $1\to 3$ splitters. We measured the ellipse phase angle for different values of the intensity ratio of ``up'' and ``down'' beams, by keeping the total power constant. The results are shown in figure \ref{cool}. The average angle  $\bar\Phi$ is very sensitive to the intensity ratio, while there is no clear evidence of any variation of the difference angle $\Delta\Phi$. From a linear fit we find that changing the up/down intensity ratio by 1\% induces a shift of $0.80\pm0.06$\,mrad on $\bar\Phi$. A parabolic fit of the $\Delta\Phi$ data provides an upper estimate of $\sim20$\,$\mu$rad$/\%^2$ to a possible small quadratic dependence. Changing the intensity ratio also modifies the position of the ellipse center and the ellipse amplitude, with sensitivity $\sim 10^{-4}$/\% and $\sim 0.5\times10^{-4}$/\%, respectively.

By recording the time of flight (TOF) of atomic clouds from launch to the detection region, we observe that changes in the up/down intensity ratio induce a vertical shift of the MOT position with a coefficient of about 0.1\,mm/\%. However, this effect cannot explain the measured shift in $\bar\Phi$; in fact, we investigated the vertical gravity gradient and the magnetic gradient in the interferometer tube (see \cite{Rosi2013} and section \ref{bpulse}): the calculated $\bar\Phi$ shift from both gradients is smaller than the measured $\sim 8$\,mrad/mm linear coefficient by one order of magnitude. We conclude that the effect of the up/down intensity ratio on $\bar\Phi$ is most probably determined from changes in the temperature of atomic clouds.

When changing instead the total power of MOT beams at constant intensity ratio, we find an upper limit of $\sim20$\,$\mu$rad/\% for the linear coefficient of both $\bar\Phi$ and $\Delta\Phi$.

\subsection{Detection}
\label{detection}

We measure the normalized number of atoms in the $F=1$ and $F=2$ states by fluorescence spectroscopy after the interferometry sequence. In the detection region, the atomic clouds cross two horizontal laser beams, resonant with the $F=2\to F'=3$ transition, having a horizontal size of $15$\,mm, a vertical size of $\sim5$\,mm and a vertical separation of $\sim20$\,mm. Both lasers are retro-reflected only for the upper $\sim3$\,mm portion of the beam, leaving the lower $\sim2$\,mm for a traveling wave which is used to blow away $F=2$ atoms right after detection. The two beams are split from a single laser source with a polarizing beam splitter (PBS) close to the detection chamber. A weak repumper beam, resonant with the $F=1\to F'=2$ transition, propagates horizontally between the two probe beams, to optically pump $F=1$ atoms before detecting them on the $F=2$ transition in the lower interaction region. The optical intensity of probe and repumper beams affects the photon scattering rate of the atoms, and thus the signal at the two detection channels. Any unbalance between the efficiencies $\eta_i$ of the two detection channels may result in principle in a shift of the ellipse phase angle. In the following we discuss possible sources of detection unbalance, and the magnitude of the effect on the gravity gradient measurement.

\subsubsection{Relative efficiency of detection channels}
\label{detection_efficiency}

The differential gravity acceleration is calculated from the phase angle of an ellipse whose points  $ (x, y)$ are given by the normalized number of atoms in the $F=1$ state for each cloud, i.e. $x=n_{11}/(n_{11}+ n_{21})$ and $y=n_{12}/(n_{12}+ n_{22})$.
However, atomic populations $n_{ij}$ are measured from the areas $A_{ij}$ of detection peaks. if the detection efficiencies of the two channels are not equal, i.e. $\xi=\eta_1/\eta_2\neq 1$, then the  Lissajous plot obtained with

\begin{equation}
\label{eggequation}
\begin{cases}x=\frac{A_{11}}{A_{11}+ A_{21}}=\frac{n_{11}}{n_{11}+\xi n_{21}}\\
y=\frac{A_{12}}{A_{12}+A_{22}}=\frac{n_{12}}{n_{12}+\xi n_{22}}\\
\end{cases}
\end{equation}

results in a distorted ellipse. The phase angle obtained from least-squares ellipse fitting depends on the relative detection efficiency $\xi$; this bias is not efficiently removed in the doubly differential scheme for $G$ measurement.  
In order to evaluate the effect of detection unbalance on noise and systematic error, we calculated the 
phase angle $\Phi$  and rms error $\delta\Phi$ of least squares ellipse fitting versus  $\xi$ using a set of synthetic data. Both  $\Phi(\xi)$ and $\delta\Phi(\xi)$ have a minimum around $\xi=1$. In order to keep the systematic error on $\Phi$ below 100\,$\mu$rad, the relative detection efficiency must be calibrated to better than 3\%. The systematic error depends on the noise level on the ellipses: in our simulations points we applied a noise level comparable to that of our typical experimental data. Another consequence of detection unbalance is a shift of the ellipse center: if $\xi$ changes, the ellipse translates along the $x=y$ direction.

An efficiency unbalance between the two detection channels may arise from either differences of size and power of probe beams, from limited repumping efficiency of $F=1$ atoms or from geometrical differences between the two optical systems for fluorescence collection.

 In principle, it is possible to compensate for any detection unbalance originated from geometrical differences by properly adjusting the intensity ratio of probe beams. However, if the probe beams have unequal intensities, the saturation parameter is different for the two detection channels; as a result, even common mode intensity fluctuations will be converted to $\xi$ changes. 

\begin{figure}
\includegraphics[width=0.5\textwidth]{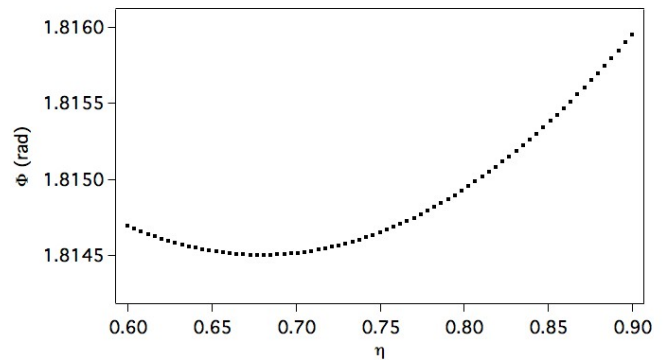}%
\caption{\label{ellipsevsdetectionefficiency}Ellipse phase angle versus detection efficiency ratio for a typical data set.}%
\end{figure}

Absolute calibration of the relative detection efficiency $\xi$ at the $\sim1\%$ level is technically challenging, due to the unavoidable differences in the geometry of collection optics. However, it is possible to determine the detection unbalance introducing $\xi$ as a parameter in eq.~(\ref{eggequation}): the value $\bar\xi$ corresponding to the minimum of  $\Phi(\xi)$ or $\delta\Phi(\xi)$ as determined from ellipse fitting represent our best estimate for the effective detection efficiency ratio. Figure \ref{ellipsevsdetectionefficiency} shows the $\Phi(\xi)$ values obtained from a set of experimental data. The location of $\bar\xi$ is the same as for the corresponding $\delta\Phi(\xi)$ curve within the experimental error. 
We checked the consistency of our method, which is equivalent to introduce an additional parameter $\xi$ in the least squares fitting, by a numerical simulation. We generated a large number of ellipses with contrast, bias, noise and detection efficiency ratio similar to our experimental conditions. We verified that our algorithm returns the correct value of $\Phi$ within $\sim100$\,$\mu$rad. In principle, the detection efficiency ratio $\xi$  might be different for the two simultaneous interferometers, due to difference in the cloud size and velocity. However, in our typical experimental conditions, we verified that minimizing $\delta\Phi$ with respect to two independent $\xi$ parameters does not change our estimate of $\Phi$ by more than $\sim100$\,$\mu$rad.

\subsubsection{Frequency fluctuations of probe laser beams}
\label{probe_frequency_fluctuations}

Frequency jitter of the detection light changes the scattering rate. During the detection sequence the scattering rate has to be constant to allow for normalization, since the interferometer ports are read out sequentially. The spectral density of frequency noise of our probe laser is $\sim 10^2$\,Hz/$\sqrt{\rm Hz}$. Given our typical values for detuning and intensity for the probe laser, and a duration of the detection sequence $\sim 15$\,ms for each cloud, the contribution of frequency noise is below the QPN limit for $10^5$ atoms.

\subsubsection{Intensity fluctuations of probe laser beams}
\label{probe_fluctuations}

In our setup, the intensity ratio of probe beams is passively stabilized to 0.1\% with a high extinction polarizer placed before the PBS. As a result, probe intensity fluctuations on time scales longer than the delay between $F=2$ and $F=1$ detection (i.e. $\sim 15$\,ms) are essentially common mode between the two channels.
However, fast fluctuations would yield noise on the measurement of normalized atomic population. 
Moreover, as seen in the previous paragraph the ellipse phase angle, bias and contrast can still depend on the total power of probe beams, as well as on the power of the repumping beam.

%
%
%

\begin{figure}
\includegraphics[width=0.5\textwidth]{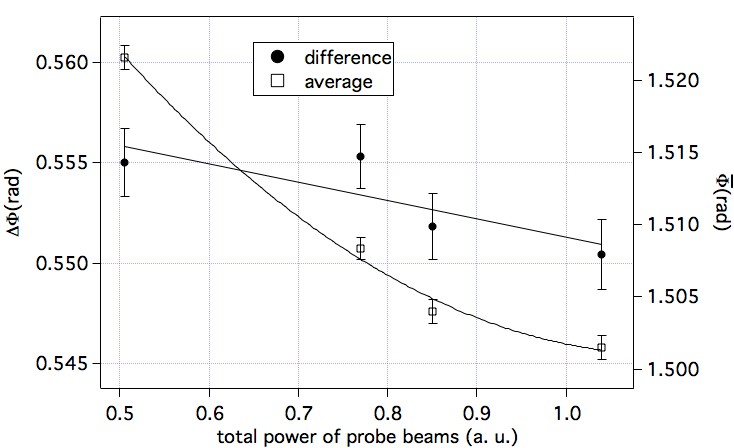}%
\caption{\label{probellipse}Average and differential ellipse angle for the two configurations of source masses, versus power of probe laser beams. Solid lines are least squares linear (black points) and parabolic (white squares) fits to the data.}%
\end{figure}

We recorded the ellipse phase angle, contrast and bias, in the two configurations of source masses, for different values of the total probe laser intensity $I_p$ and of the intensity ratio between the two probe beams.
 In both cases, 
 the change in detection efficiency ratio produces a translation and a distortion of the ellipses, which modify the center, amplitude and rotation angle of the best fitting ellipse. 
As an example, the plot of ellipse phase angle versus total intensity $I_p$ is shown in figure \ref{probellipse}. 
The 
slope of the $\bar\Phi(I_p)$ curve decreases when $I_p$ is above the saturation intensity. Around our typical experimental conditions, $I_p\cdot\partial \bar\Phi/\partial I_p= -0.15\pm 0.04$\,mrad/\%. For the sensitivity of the difference angle $\Delta\Phi$ on $I_p$ we derive an upper limit of $\sim90$\,$\mu$rad/\%. The sensitivity of $\bar\Phi$ on intensity ratio is $\sim40$\,$\mu$rad/\%, while the sensitivity of $\Delta\Phi$ is lower than $\sim40$\,$\mu$rad/\%.

The $X$ and $Y$ coordinates of ellipse center depend on the total probe intensity at fixed ratio with a sensitivity $I_P\cdot\partial B/\partial I_p\sim I_P\cdot\partial D/\partial I_p\sim -4\times10^{-4}$\,/\%. The ellipse amplitude has a weaker sensitivity $I_P\cdot\partial A/\partial I_p\sim I_P\cdot\partial C/\partial I_p\sim -0.6\times10^{-4}$\,/\%. The sensitivities of ellipse center and amplitude on intensity ratio are $(1.6\pm0.2)\times10^{-3}$\,/\% and  $(0.8\pm0.1)\times10^{-4}$\,/\%, respectively.

\begin{figure}
\includegraphics[width=0.5\textwidth]{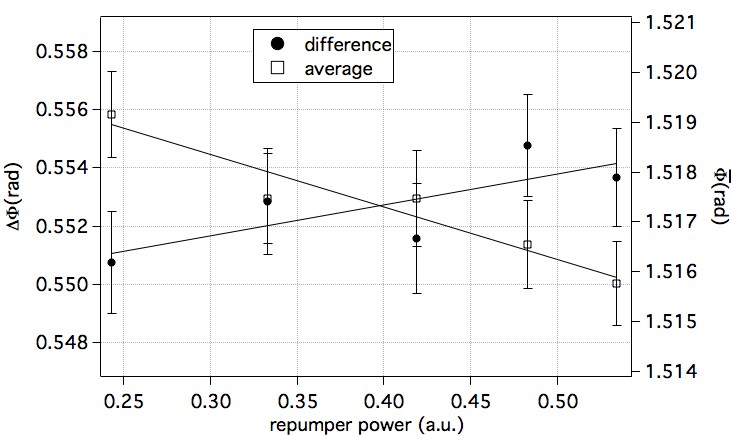}%
\caption{\label{repumper}Average and differential ellipse angle for the two configuration of source masses, versus power of repumping light in the probe beam. Solid lines are least squares linear fits to the data.}%
\end{figure}

We also measured the effect of intensity changes in the repumper beam. The results are shown in figure \ref{repumper}. Changes in the optical intensity of repumper $I_r$ modify the detection efficiency in the $F=1$ channel; around our typical experimental conditions, the phase angle $\bar\Phi(I_r)$ decreases with repumper power with a slope $I_r\cdot\partial \bar\Phi/\partial I_r= -0.10\pm0.02$\,mrad/\%. The sensitivity of the differential phase angle  $\Delta\Phi$  is below 0.1\,mrad/\%, while ellipse center and amplitude have sensitivities of $(-1.5\pm0.1)\times10^{-4}$\,/\% and  $(-0.7\pm0.1)\times10^{-4}$\,/\%, respectively.

\subsubsection{Noise and biases arising from detection of different atomic velocity classes}
\label{VS}

\begin{figure}
\includegraphics[width=0.5\textwidth]{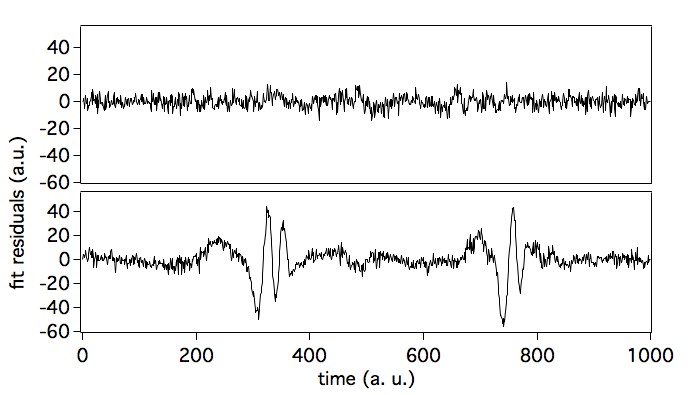}%
\caption{\label{fitresiduals}Fit residuals of detection peaks after a single velocity selection. Upper plot: $F=2$ state; lower plot: $F=1$ state.}%
\end{figure}

Shortly before the atom interferometry sequence, we select a narrow class of vertical velocity from the thermal clouds; 
an efficient elimination of the thermal background from the velocity selected atoms is important to achieve high contrast ellipses and to control systematic  shifts on the ellipse phase angle. In order to investigate the effect of residual thermal atoms on the measurement accuracy, we compared two different methods for velocity selection. 
After launch in the atomic fountain, almost all the atoms in the thermal cloud are pumped 
in the $F=2$ state. 
In the first method (single pulse selection), we apply a velocity selective Raman pulse, tuned to the
 $|F = 2, m_F = 0 \rangle \to | F = 1, m_F = 0\rangle$ transition, shortly after launch. Atoms in a narrow velocity class, which is Doppler shifted to resonance, are pumped into the $| F = 1, m_F = 0\rangle$ state. The vertical velocity spread $\delta v$ of selected atoms is determined by the duration $\tau$ of the Raman pulse, i.e. $\delta v\simeq (\tau k_{e})^{-1}\sim 1.3$\,mm/s, corresponding to a temperature of $\sim 18$\,nK. 
Before starting the interferometry sequence, we then eliminate the residual atoms in the $F = 2$ state
with a 5\,ms pulse (slightly divergent, circularly polarized) tuned to the cycling $|F=2\rangle\to |F=3\rangle$ transition. 
However, when using this method for state and velocity selection we always find a non negligible fraction of thermal atoms in the $F=1$ detection, 
producing a wide pedestal below the detection peak. Figure \ref{fitresiduals} shows typical fit residuals for $F=2$ and $F=1$ atoms (see section \ref{detectionnoise} about peak fitting models). The thermal pedestal is indeed resulting from the off-resonant photon scattering from the Raman beams during the velocity selection pulse.
In the presence of a large fraction of thermal atoms, determination of the $F=1$ peak area  and  calculation of the normalized $F=1$ population are not reliable. This is due to both the large RMS error of the peak fitting, and to the fact that part of the thermal atoms do not interact with Raman lasers in the interferometry sequence, while still being detected.

The geometry of our apparatus prevents the possibility to employ Zeeman state selection with microwave pumping. In order to eliminate the thermal pedestal in the $F=1$ channel, we implemented a different state and velocity selection (triple pulse selection), based on the application of three Raman pulses to transfer the atoms back and forth between the $|F = 2, m_F = 0 \rangle $ and $ | F = 1, m_F = 0\rangle$ states. After each Raman pulse, we apply a resonant laser pulse to blow away the remaining atoms in the initial state. The blow-away laser for $F=1$ atoms is resonant to the $|F=1\rangle\to |F=0\rangle$ transition. After the application of the triple pulse velocity selection, the $F=1$ peaks show no detectable thermal pedestal, and no clear structures in the fit residuals. The RMS of fit residuals is now the same for the two channels (see section  \ref{detectionnoise}). 
With the number of thermal atoms detected in the pedestal of TOF signals reduced by a factor  $>30$, the systematic effects on the gravity gradient measurements can be controlled to better than 100\,$\mu$rad.

A drawback of the triple pulse velocity selection is a reduction in the number of selected atoms by a factor $\sim2$, because of the limited ($\sim70$\,\%) efficiency of Raman $\pi$ pulses. However, the reduced signal is well compensated by a larger contrast of the interference fringes. As a result, the sensitivity of gravity gradient measurement is improved with respect to the use of single pulse velocity selection (see section \ref{short}).

\subsection{Influence of Raman lasers on noise and bias of the atom interferometer}
\label{Raman}

Fluctuations in the frequency, intensity and alignment of Raman laser beams may induce changes in the ellipse phase angle. One of the two Raman lasers (master) is frequency locked, with a red detuning of 2\,GHz, to the reference laser, which is frequency stabilized to the $|F=2\rangle\to |F=3\rangle$ $^{87}$Rb line with the modulation transfer spectroscopy technique \cite{Bjorklund83}. The absolute frequency of the master laser is stable within 0.5\,MHz. The other Raman laser (slave) is phase locked to the master laser, with an offset of 6.8 \,GHz given by a RF synthesizer.  In our experimental conditions the effect of  phase and frequency fluctuations of Raman lasers on the ellipse phase angle is negligible. 

When the detuning of the Raman lasers is fixed, the intensities of Raman beams determine the Rabi frequency, i.e. the probability of the Raman transitions, as well as its spatial distribution through the inhomogeneous light shift. We set the ratio between the optical intensity of master and slave lasers, $I_M$ and $I_S$, to the value which cancels the first-order light-shift at the frequency detuning from the $|F=2\rangle\to |F=3\rangle$ resonance selected for the Raman lasers (see next section). We fix the duration of Raman pulses, and adjust the total optical power of the Raman beams in order to maximize the efficiency of $\pi$ pulses. Intensity fluctuations or drifts may change the ellipse contrast; more importantly, they might change the velocity class of selected atoms because of residual light shift.

\subsubsection{Effect of light shift}
\label{lightshift}

To precisely cancel the first order light shift, we measure the vertical velocity of the atomic clouds after a single $\pi$ pulse versus the power of Raman beams. Velocity changes are detected with the time of flight method, i.e. by measuring the arrival time of the atomic clouds in the detection region. 

\begin{figure}
\includegraphics[width=0.5\textwidth]{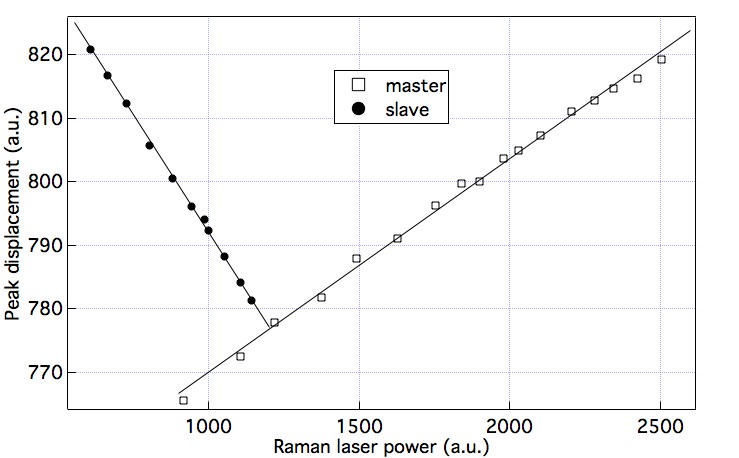}%
\caption{\label{lightshift_measure} Vertical displacement of velocity selected clouds versus power of master and slave Raman beams. Solid lines are least squares linear fits to the data.}%
\end{figure}

At fixed detuning of the Raman lasers, the resonant frequency of the Raman transition can be written as

\begin{equation} 
f(I_M,I_S)=f_0+C_M I_M + C_S I_S +O(I^2)
\end{equation}
 where $f_0$ is the unperturbed resonance, and $I_M$ and $I_S$ are the intensities of master and slave laser beams, respectively. 
For a given value of $I_M$, we measured the position of the upper cloud versus $I_S$, and vice versa. Figure \ref{lightshift_measure} shows the results. We determine the $C_M$ and $C_S$ coefficients by a linear fit to experimental data. By properly setting the intensity ratio 

\begin{equation} 
\frac{I_M}{I_S}=-\frac{C_S}{C_M}
\end{equation}
we can cancel the first order light shift independently on the total Raman power. We determine this optimal ratio to be $C_S/C_M=-0.45\pm0.02$.

\subsubsection{Intensity fluctuations of Raman lasers}
\label{Ramanfluctuations}

\begin{figure}
 \includegraphics[width=0.5\textwidth]{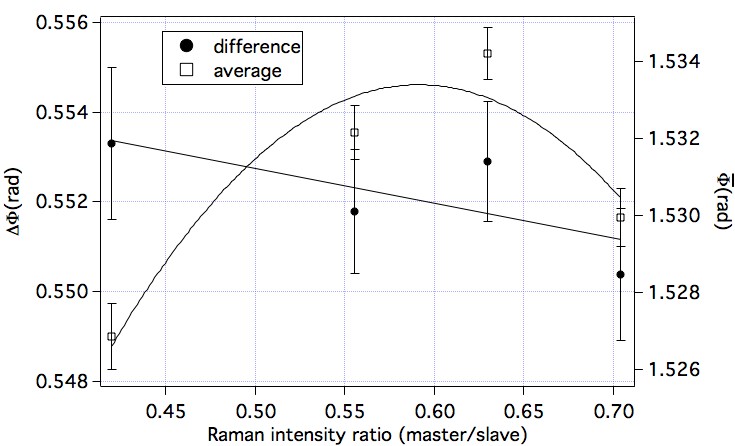}%
\caption{\label{ramanratio}Average and differential ellipse angle for the two configuration of source masses, versus intensity ratio of Raman lasers. Solid lines are least squares linear (black points) and parabolic (white squares) fits to the data.}%
\end{figure}

In order to estimate the effect of Raman laser intensities on the gravity gradient measurement, we recorded the ellipse phase angle for different values of the total Raman intensity $I_{M}+I_S$ and of the intensity ratio $I_M/I_S$ in the two configurations of source masses. 
The behavior of $\bar\Phi$ and $\Delta\Phi$ versus the intensity ratio of Raman laser beams is shown in figure \ref{ramanratio}. The shift of $\bar\Phi$ is maximum for an intensity ratio around $0.59\pm0.05$.
 From a parabolic fit we determine a curvature of $20\pm4$\,$\mu$rad/\%$^2$ around the maximum. From the $\Delta\Phi$ plot we extract  a limit of 0.1\,$\mu$rad/\% for the sensitivity of the differential angle on the Raman intensity ratio.

In a similar way we determine the influence of total Raman intensity when $I_M/I_S=0.45$.  We measure a $\bar\Phi$ sensitivity of $0.30\pm0.04$\,mrad/\%, which can be attributed to the combination of residual first-order light shift and second order light shift.  The sensitivity of  $\Delta\Phi$ on total intensity ratio is below 0.1\,$\mu$rad/\%.

Changing the intensity ratio and overall intensity of Raman beams also modifies the position of the ellipse center, with sensitivity $\sim 10^{-4}$/\%, and the ellipse amplitude,  with sensitivity $\sim 0.5\times10^{-4}$/\%.

\subsubsection{Alignment fluctuations of Raman beams}
\label{Ramantilt}

We align the Raman beams along the vertical direction with sub-mrad precision with
   the aid of a liquid mirror. However, small fluctuation in the propagation direction of  Raman beams would couple with the gravity gradient measurement through the Coriolis effect.

\begin{figure}
 \includegraphics[width=0.5\textwidth]{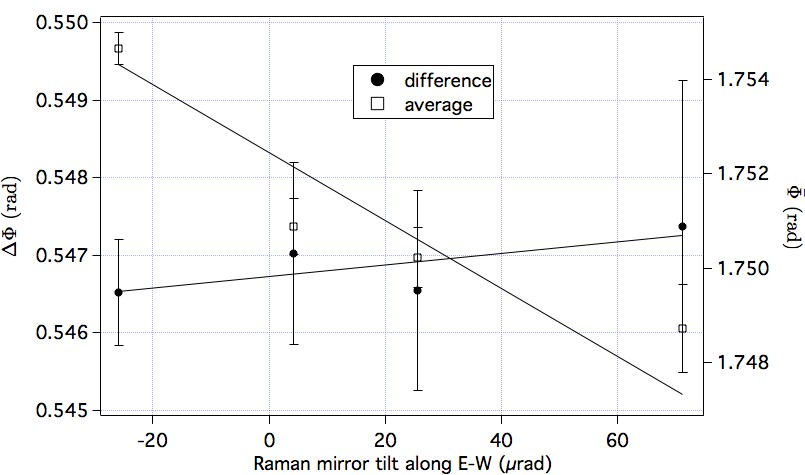}%
\caption{\label{tiltmeter}Average and differential ellipse angle for the two configuration of source masses, versus tilt of the Raman mirror along the $y$ direction. Solid lines are linear fits to the data.}%
\end{figure}

Assuming a small inclination $\theta$ of the $k_{e}$ vector along the East-West direction, and  assuming that the upper and lower atomic clouds are launched vertically with initial velocities $v_u\simeq4.3$\,m/s and $v_l\simeq3.5$\,m/s respectively, the resulting shift of the ellipse phase angle due to first order Coriolis force is 

\begin{equation}
\label{Corioliseq}
\phi_{C}=-2\Omega_E k_{e}T^2(v_u-v_l)\cos{\alpha_l}\sin{\theta}
\end{equation}
where $\Omega_E$ is the Earth's rotation rate and $\alpha_l$ is the latitude angle at the location of our laboratory. In our case, with $T=160$\,ms and  $\alpha_l\simeq43^{\circ}$, the Coriolis shift is $ \phi_{C} \simeq-34\theta$. All of the mountings for the optics delivering the Raman beams are chosen to be extremely rigid, and fluctuations in the propagation direction are essentially dominated by the tilt of the retro-reflection mirror which is mounted on the top of the structure holding the source masses. We directly observed  the effect of mirror tilt on the ellipse phase angle. The Raman retro-flection mirror is mounted on a precision, dual-axis tiltmeter, that measures the inclination $\theta_x$ and $\theta_y$ along two axes. The  $y$ axis is oriented along the West-East direction within a few degrees. We recorded several ellipses for different values of  the mirror tilt  $\theta_y$, by keeping $\theta_x$  constant, and vice versa. The results of average and differential ellipse phase angle versus $\theta_y$ are shown in figure \ref{tiltmeter}.

From a linear fit to the $\bar\Phi$ data, we derive a sensitivity of $-37\pm5$\,mrad/mrad, in good agreement with eq.~(\ref{Corioliseq}). In a similar way, we measure a sensitivity of $-5\pm2$\,mrad/mrad for the  $\bar\Phi$  dependence on  $\theta_x$, which is compatible with eq.~(\ref{Corioliseq}) assuming an angle of $\sim8^{\circ}$ between the $x$ axis and the North-South direction.

From the $\Delta\Phi$ data, there is no evidence of any direct effect of mirror tilt on the differential phase angle. Nevertheless, the influence of Coriolis shift on $G$ measurement is not negligible, because of a tiny deformation of the mechanical structure which is induced by source masses. In fact we observe that the vertical translation of source masses induces a tilt of the Raman mirror, so that $\theta_y$ changes by $\sim10$\,$\mu$rad between the two configurations. This results in a bias of $\sim350$\,$\mu$rad on $\Delta\Phi$, corresponding to a systematic error of $6\times10^{-4}$ on $G$. This bias can be easily reduced by either correcting the $\Delta\Phi$ data for the measured mirror tilt in post-processing, or by actively stabilizing the angle of the mirror with PZT actuators.

\subsubsection{Earth's rotation compensation}
\label{Coriolis}

As long as the atoms are launched with some residual horizontal velocity along the East-West direction, the Coriolis force yields a phase shift on the atom interferometer output. This represents a source of both systematic errors  and noise. The systematic error on the gravity gradient measurement is proportional to the East-West component of the average velocity difference between the two atomic clouds (see eq.~(\ref{Corioliseq})). Such effect would cancel out in the doubly differential 
measurement of local masses, provided that the atomic velocities do not change when moving the source masses.
According to eq.~(\ref{Corioliseq}), a change $\Delta\theta$ in the launching direction of the atomic fountain would produce a change $\Delta\phi\sim-34\Delta\theta$ in the differential ellipse angle. In order to keep the systematic effect on $G$ measurement within $\sim50\,$ppm, i.e. $\Delta\phi_{C}<30\,\mu$rad, it is necessary to control possible changes  $\Delta\theta$ in the launching direction within $\sim1$\,$\mu$rad, i.e. to measure the shift in the center of atomic distribution with micrometer precision. 

\begin{figure}
 \includegraphics[width=0.5\textwidth]{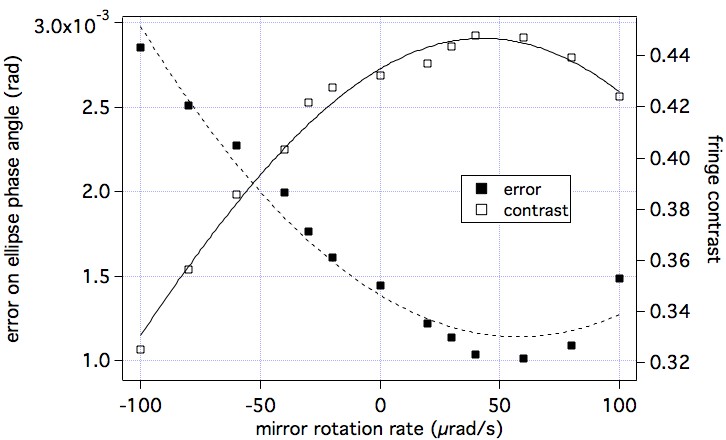}%
\caption{\label{coriolisplot}Effect of Raman mirror rotation on the atom interferometer sensitivity. The plot shows the fit error on ellipse phase angle and the fringe contrast versus the Raman mirror rotation rate along the North-South direction. The fringe contrast is defined as $2A$ with reference to eq.~(\ref{ellipseq}). Lines are parabolic fits to experimental data.}%
\end{figure}

On the other hand, the horizontal velocity spread corresponding to the $\sim3$\,$\mu$K transverse atomic temperature is expected to contribute to the noise on the ellipse phase angle via Coriolis effect. This is shown in figure \ref{coriolisplot}; we apply a uniform rotation rate to the retro-reflecting Raman mirror during the atom interferometry sequence by means of PZT actuators, as suggested in \cite{Hogan09,Lan12}. Figure \ref{coriolisplot} shows the rms error of ellipse phase angle versus the mirror rotation rate; the rotation axis is roughly oriented along the North-South direction. The optimal rotation rate, corresponding to the maximum contrast, is equivalent to the opposite of the local projection of the Earth rotation rate on the horizontal plane. In such conditions, the rms noise on ellipse fitting is minimum. 
As a result, the compensated error on ellipse angle is  $\sim50$\,\% lower than without compensation, while the contrast increases by $\sim4$\,\% only.

\subsection{Effect of magnetic fields}
\label{Bfield}

Magnetic fields affect the atom interferometry measurement mainly in two ways: through the impact on atomic trajectories, and through the Zeeman shift of energy levels along the cloud's trajectories.  We use several coils to separately create well controlled bias fields in the MOT region and in the fountain tube. In our experiment, the interferometer tube is surrounded by  two concentric cylindrical $\mu$-metal layers, that attenuate external magnetic fields by more than 60\,dB in the region of the atom interferometry sequence. The MOT and detection chambers, on the contrary, are not shielded.

\subsubsection{MOT compensation coils}
\label{shimfield}

\begin{figure}
\includegraphics[width=0.5\textwidth]{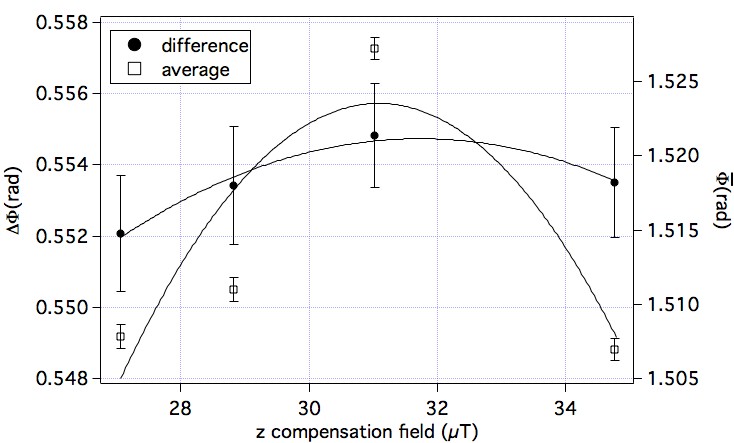}%
\caption{\label{shim}Average and differential ellipse angle for the two configuration of source masses, versus current in vertical compensation coils. Solid lines are parabolic fit to experimental data: however the black points with their error bars are also consistent 
  with a constant.}%
\end{figure}

The launching direction of the atoms in the fountain is sensitive to the magnetic field in the MOT region. In fact, we perform a fine tuning of the fountain alignment by acting on the current of three pairs of Helmholtz coils, which are oriented along three orthogonal axes to create a uniform bias field at the  position of the MOT. In order to investigate the sensitivity of the gravity gradient measurements on the magnetic fields in the MOT region, we recorded several ellipses for different values of  the current in the compensation coils. As an example, figure \ref{shim} shows the plot of average and difference ellipse angle for the two configurations of source masses, versus the bias field produced by the vertical compensation coils. The  $\bar\Phi$ data clearly show the presence of maximum around 31\,$\mu$T, with a curvature of $1.16\pm0.06$\,mrad/$\mu$T$^2$.

At our typical operating conditions (i.e. around 29\,$\mu$T) the linear sensitivity is $\sim4.5$\,mrad/$\mu$T. Since the vertical compensation coils produce a field of 0.22\,mT/A, this converts into a sensitivity of $\sim1$\,mrad/mA. 

Again we find no evidence of any effect on the differential ellipse angle. 

\subsubsection{Bias field in the interferometer tube}
\label{bpulse}

A 1\,m long solenoid inside the $\mu$-metal tube creates a uniform magnetic field $B_0\simeq29$\,$\mu$T to define the quantization axis during the atom interferometry sequence. The interferometer sequence is applied to atoms in the $m_F=0$ 
state; a uniform magnetic field produces a constant energy shift, yielding no extra phase shift due to the symmetry of the interferometer. However, a magnetic gradient would induce  a phase shift 
on each of the two interferometers through
the second order Zeeman effect. The magnitude of the phase shift depends on the initial velocity $v=gT+gt_a$ at the first $\frac{\pi}{2}$-pulse, where $t_a$ is the
    time mismatch between the time at which the unperturbed cloud reaches 
    the apogee and the $\pi$-pulse (see section \ref{apparatus}). We typically use $t_a\simeq5$\,ms. Assuming a linear magnetic gradient $\gamma$, the differential phase shift in the gravity gradiometer is

\begin{equation}
\label{gradeq2}
\delta\Phi_{\gamma}=\pi\alpha \gamma^2 (v_r+2 g t_a )T^2 \Delta z
\end{equation}
where $\alpha\simeq57.5$\,GHz/T$^2$ is the differential coefficient of quadratic Zeeman shift, $v_r$ is the recoil velocity and $\Delta z$ is the vertical separation of the atomic clouds. 

\begin{figure}
\includegraphics[width=0.5\textwidth]{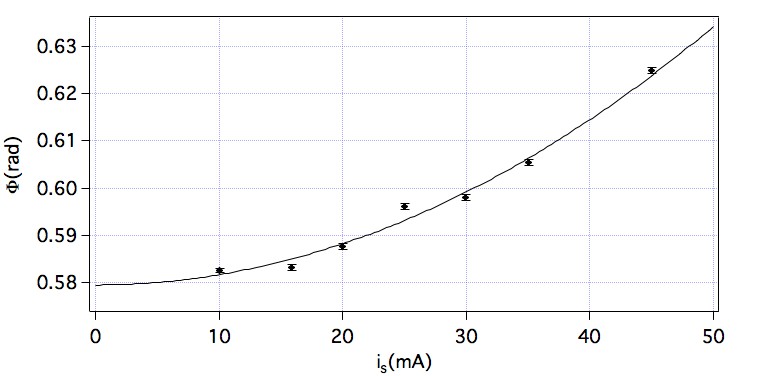}%
\caption{\label{biasnopulse}Ellipse phase angle versus current in vertical bias solenoid, without magnetic pulse from the short coil. The solid line is a linear fit to experimental data.}%
\end{figure}

We investigated the presence of magnetic gradients by recording the ellipse phase angle versus the current $i_s$ in the solenoid. The results are shown in figure \ref{biasnopulse}. The data are reasonably consistent with a parabola with the vertex at 
$i_s=0$ and curvature $22.0\pm0.2$\,$\mu$rad/mA$^2$; since the solenoid produces a field $\frac{\partial B_0}{\partial i_s}\sim 1.445$\,mT/A, this corresponds to $\sim7$\,$\mu$rad/$\mu$T$^2$.
The bias coil does not generate a perfectly uniform field: the magnetic gradient is proportional to the current  $i_s$ in the solenoid, yielding the quadratic scaling. 
In an ideal solenoid the magnetic field would have a parabolic shape, and the theoretical differential phase shift $\delta\Phi_{\gamma}$ would have a quadratic dependence of
     the order of $\sim5$\,mrad/mA$^2$, which cannot explain the observed dependence in figure \ref{biasnopulse}. We assume that the solenoid is not ideal
     and derive from eq.~(\ref{gradeq2}) an estimate for the linear gradient in the solenoid; at our typical working point, i.e. 20 mA, we obtain
     $\gamma\simeq8$\,$\mu$T/m. On the other hand, there is no detectable stray magnetic field in the tube: an upper 
limit is obtained by fitting the data with a parabola with a linear term. 
The vertex is at $i_s=2$\,$\mu$A which corresponds to 3\,nT.
In our typical working conditions, i.e. with $i_s\simeq20$\,mA, the sensitivity to $i_s$ is below 0.9\,mrad/mA. 

For the measurement of gravity gradient, it is necessary to extrapolate the angle at 
$i_s=0$. We obtain an angle of $580\pm1$\,mrad; after correcting for the gravitational effect of closest masses \cite{Sorrentino12}, we obtain a value of $(3.135\pm0.007)\times 10^{-6}$\,s$^{-2}$ for the gravity gradient.

For the 
measurement of local source masses, a short (20\,cm) coil creates a square pulse of length $\tau\sim10$\,ms
of magnetic field, around the apogee of the lower atomic cloud, during the second half of the  interferometry sequence. During the pulse a field difference  $\Delta B\sim10$\,$\mu$T is induced
    between the two clouds.  The time $\tau$ is so short that the clouds do not move 
    by a distance over which $\Delta B$ changes significantly. The corresponding extra phase shift
\begin{equation}
\label{gradeq3}
\delta\Phi_B=2\pi\alpha \tau [(B_0+\Delta B)^2-B_0^2]
\end{equation}
 is used to make the eccentricity of ellipses low enough and symmetric in the two configurations of source masses.

\begin{figure}
\includegraphics[width=0.5\textwidth]{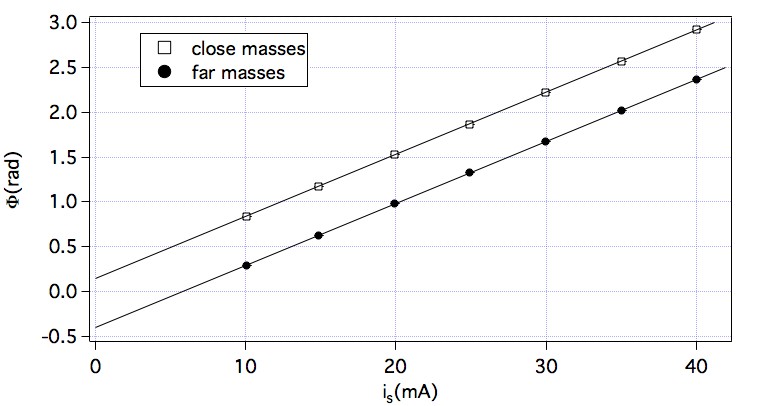}%
\caption{\label{biaspulse}Ellipse phase angle for the two configurations of source masses versus current in vertical bias solenoid, with applied magnetic pulse from the short coil (see text). Error bars are not visible on this scale. Solid line are linear fits to experimental data.}%
\end{figure}

The magnetic gradient induced by the short coil strongly enhances the sensitivity of ellipse phase angle to solenoid bias current. This fact can be used to detect possible effects of source masses on the static magnetic field in the interferometer tube. 
This idea is illustrated in figure \ref{biaspulse}, where the ellipse phase angle for the two configurations of source masses is plotted versus $i_s$. The two plots show a linear dependence, in agreement with eq.~(\ref{gradeq3}) which predicts a slope $4\pi\alpha \tau\Delta B\frac{\partial B_0}{\partial i_s}\simeq72$\,mrad/mA. The measured slope $\partial \Phi/\partial i_s=69\pm1$\,mrad/mA is the same, within the experimental uncertainty of 0.2\%, for the two configurations of source masses, i.e. we found no evidence of any influence of source masses. Moreover, possible systematic errors in the gravity gradient measurement from second order Zeeman effect are efficiently removed through the $k$-reversal technique (see section \ref{long}).

\section{High precision measurement of 
differential gravity}
\label{measurements}

The data presented in section \ref{parameters} allow to identify the main limits to the stability of $\Phi$ measurements, once the typical fluctuations of the parameters are known. We constantly monitor the value of most relevant experimental parameters: the power of MOT, probe, repumper and Raman laser beams, the current in MOT compensation coils, in pulse coil and in bias solenoid, the tilt of Raman mirror, as well as the temperature in different points of the apparatus with a high resolution data logger. In the following, we show how the active control of such parameters allows to improve the precision on gravity gradient and $G$ measurements.

\subsection{Active control of main experimental parameters}
\label{Servos}

Table \ref{phisensitivity} 
summarizes the results of our characterization measurements about the influence of most relevant parameters on  average and differential ellipse phase angle, respectively. The last two columns give the typical RMS fluctuations of the parameters on two relevant time scales, i.e. over $t_{e}\sim0.5$\,hr and over one day, respectively. The impact on long term stability of the gradient and $G$ measurements are discussed in section \ref{long}.

Table \ref{phisensitivity} shows that the main contributions will arise from instability of MOT laser beams intensity ratio, probe beams total power, current in the bias solenoid and MOT compensation coils, and tilt of the Raman mirror. However, noise in the coils current is fairly white, and would not entail the long term stability, while fluctuations in laser powers and mirror tilt exhibit a low frequency flickering.

In order to improve the long-term stability, we actively stabilize the main experimental parameters, i.e. the optical intensity of cooling, Raman and probe laser beams, acting
on the RF power driving acousto-optical modulators, and the Raman mirror tilt, acting on the piezo tip/tilt system. 

The servo on cooling and Raman lasers intensity, as well as on Raman mirror tilt, is implemented by means of a slow digital loop: we sample the four powers (up and down cooling beams, master and slave Raman beams) and the two components of mirror tilt every 72 experimental cycles (about 2 minutes); then we drive the RF power of the 
corresponding AOMs, and the PZTs on Raman mirror, through a numerical loop filter. Residual fluctuations are below $0.3$\,\%. The servo on probe laser intensity is implemented by means of a fast analog loop, sampling the laser intensity with a photodiode and acting on the RF power of the 
corresponding AOM.

\begin{widetext}

\begin{table}
\caption{\label{phisensitivity}Sensitivity of average and differential phase angle, contrast and bias of ellipses to most relevant parameters.}
\begin{tabular}{lllllll}
\hline\noalign{\smallskip}
Parameter $\alpha$ & $\bar\Phi(\alpha)$ slope  & $\Delta\Phi(\alpha)$ slope  & Contrast sensitivity & Bias sensitivity &  $\sqrt{\langle\delta\alpha\rangle^2_{t_{e}}}$ & $\sqrt{\langle\delta\alpha\rangle^2_{1 day}}$ \\
\noalign{\smallskip}\hline\noalign{\smallskip}
Probe power ratio & 40\,$\mu$rad/\% & $< 40$\,$\mu$rad/\% & $0.8\times10^{-4}$\,/\% & $-1.6\times10^{-3}$\,/\%	& 0.1\%  & 0.1\% \\
Probe power &  $-0.15\pm0.04$\,mrad/\% &  $<0.09$\,mrad/\% & $-0.6\times10^{-4}$\,/\% & $-4\times10^{-4}$\,/\% & 0.5\% & 2\% \\ 
Repumper power & $-0.10\pm0.02$\,mrad/\% &    $<0.01$\,mrad/\% & $-0.7\times10^{-4}$\,/\% & $-1.5\times10^{-4}$\,/\% & 0.5\% & 2\%\\
Raman intensity ratio & $20\pm4$\,$\mu$rad/\%$^2$ &   $<0.1$\,mrad/\%  & $0.5\times10^{-4}$/\% & $1\times10^{-4}$/\% & 0.5\% &  2\%\\
Raman total intensity & $0.30\pm0.04$\,mrad/\% &     $<0.1$\,mrad/\%  & $0.5\times10^{-4}$/\% & $1\times10^{-4}$/\% & 0.5\% &  2\%\\
MOT total power & $<20$\,$\mu$rad/\% &    $<20$\,$\mu$rad/\%   & $<0.5\times10^{-4}$/\% & $<1\times10^{-4}$/\% & 0.5\%  &  2\%\\
MOT power ratio & $0.80\pm0.06$\,mrad/\% &    $20\pm10$\,$\mu$rad/\%$^2$  & $0.5\times10^{-4}$\,/\% & $1\times10^{-4}$\,/\% & 0.5\% &  2\%\\
vert. MOT comp. coil  & $56\pm3$\,$\mu$rad/mA$^2$  & $<10$\,$\mu$rad/mA & $<0.5\times10^{-4}/$mA  & $<1\times10^{-4}/$mA & 10$\mu$A & 20\,$\mu$A \\
bias solenoid (no pulse) & $22\pm2$\,$\mu$rad/mA$^2$  & n.a. & n.a. & n.a. &  10$\mu$A  &  20\,$\mu$A\\
bias solenoid (with pulse) & $69\pm1$\,mrad/mA  & $<20$\,$\mu$rad/mA$^2$ & $<0.5\times10^{-4}/$mA & $<1\times10^{-4}/$mA  &  10$\mu$A  &   20$\mu$A\\
Raman mirror E-W tilt & $37\pm5$\,mrad/mrad  &  $<5$\,mrad/mrad   & $<1\times10^{-3}/$mrad & $<1\times10^{-3}/$mrad & 1\,$\mu$rad &   10\,$\mu$rad\\
\noalign{\smallskip}\hline
\end{tabular}
\end{table}

\end{widetext}

\subsection{Sensitivity}
\label{short}

\begin{figure}
\includegraphics[width=0.5\textwidth]{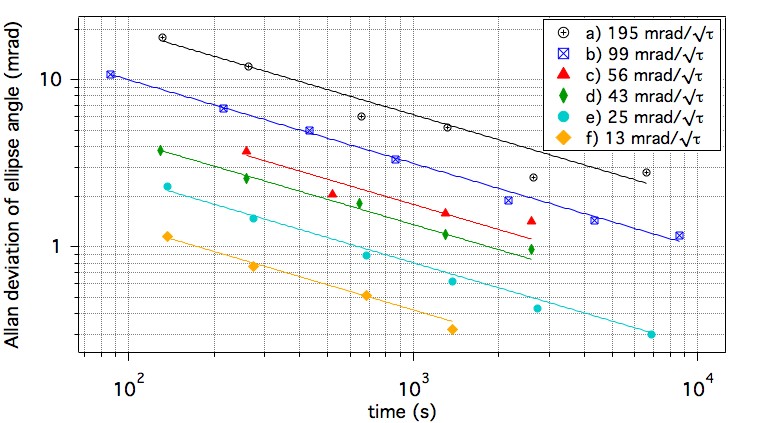}%
\caption{\label{allancomparison} (Color online) Allan deviation of the ellipse phase angle in different configurations of the experiment. Data in a) correspond to the experiment status described in \cite{Lamporesi08}; data in b) correspond to the experiment status described in \cite{Sorrentino10}, where a larger number of atoms and a faster repetition rate resulted from the implementation of a 2D-MOT and more powerful Raman laser sources;  c) resulted after minimizing the stray light at detection photodiodes; in d) we improved the contrast by implementing the triple-pulse velocity selection, and we reduced the technical noise on photodiodes with an improved readout electronics; in e) we further improved the number of atoms and applied the active stabilization of cooling, detection and Raman laser beams intensity, and of the Raman mirror tilt; in f) Earth rotation was compensated with a piezo-driven tip tilt mirror.}%
\end{figure}

In order to evaluate the sensitivity of our gradiometer, we split the atom interferometer data into groups of 72 consecutive points, and obtain a value for $\Phi$  with its estimated error from each group by ellipse fitting. 
We then  evaluate the Allan deviation of $\Phi$.
Figure \ref{allancomparison} shows the Allan deviation of ellipse phase angle in different conditions. Several improvements of the apparatus have allowed to increase the number of atoms and the repetition rate of the experiment, and also to reduce the technical noise at detection and increase the ellipse contrast. 
We currently achieve a sensitivity
of 13\,mrad at 1\,s, 
in agreement with the calculated QPN limit for $2\times 10^5$ atoms, and corresponding
to a sensitivity to differential accelerations of  $3\times10^{-9}$\,$g$ at 1\,s,  about a factor seven better than in \cite{Sorrentino10}.
We can estimate the contribution of contrast and center fluctuations from the observed sensitivity to the most relevant experimental parameters, as obtained with the same method as for the $\Phi$ sensitivity described in section \ref{parameters}, and from the typical fluctuations of such parameters on the time scale of $t_{e}$. We find that $\delta A$ and $\delta B$ are smaller than $\delta x_{d}$, which is in agreement with the fact that the observed sensitivity is close to the QPN limit. Also noise  $\delta\Delta\Phi$  on the differential phase appears to be negligible at this stage.

\subsection{Reproducibility and long term stability}
\label{long}

\begin{figure}
\includegraphics[width=0.45\textwidth]{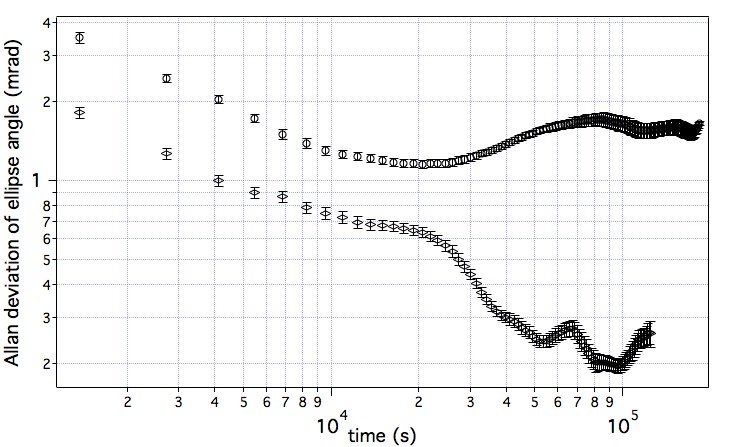}%
\caption{\label{allan_stab} Allan deviation plots of the ellipse phase angle in two different conditions; (upper points) without active stabilization of main experimental parameters; (lower points) with active intensity stabilization of cooling and detection lasers, and Coriolis compensation.}%
\end{figure}

As a first test of the long term stability of our apparatus, we observe the  statistical fluctuations of the gradiometer measurements over about 20 hours, keeping the source masses in a fixed position, and without active stabilizations of laser intensities and Raman mirror tilt. At the same time we monitor the value of most relevant experimental parameters: the power of MOT, probe, repumper and Raman laser beams, the current in MOT compensation coils, in pulse coil and in bias solenoid, the tilt of Raman mirror, as well as the temperature in different points of the apparatus.
Figure \ref{allan_stab} shows a typical Allan deviation plot for a  20\,hrs long measurement. For integration times $\tau$ lower than $\sim30$\,min the  Allan deviation scales as the inverse of the square root of $\tau$. For longer times we observe a bump, indicating a slow fluctuation of  $\phi$  with a period of a few hours. The $\Phi$ data are well correlated with the measured temperature of the laboratory. All the laser powers, as well as the Raman mirror tilt, are well correlated with the temperature with absolute values of the correlation coefficients ranging from $\sim0.7\div0.9$. 

%
%

We then tested the effect of active stabilization of the main experimental parameters. The results are shown in figure \ref{allan_stab}; the active control of cooling, Raman and probe laser intensities, together with Coriolis compensation, considerably improves both the short and long term stability. We reach a resolution of $\sim0.2$\,mrad on the ellipse phase angle, corresponding to $\sim5\times10^{-11}$\,g after an integration time of about two hours.

\begin{figure}
\includegraphics[width=0.45\textwidth]{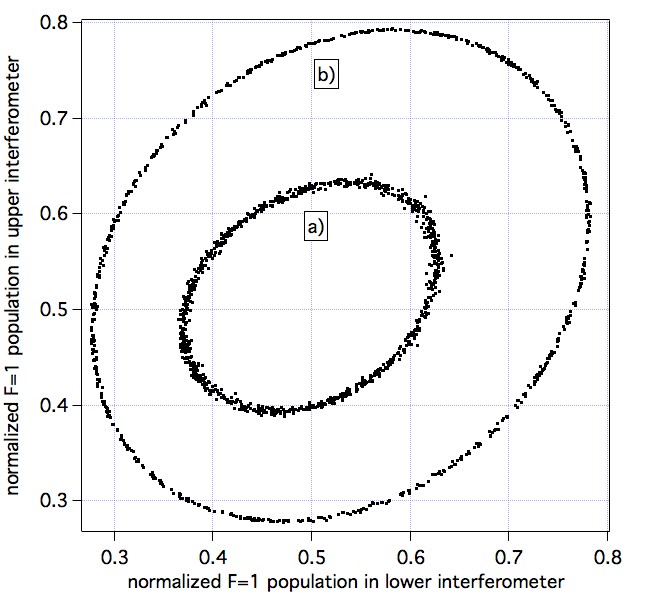}%
\caption{\label{ellipses_comparison} Elliptical plots for configuration $C_2$ of the source masses; data in a) correspond to the experiment status described in \cite{Sorrentino10}; data in b) resulted after reduction of technical detection noise, active intensity stabilization of cooling, Raman and detection lasers, and Coriolis compensation.}%
\end{figure}

%
%

We  tested the long term stability of the measurement of the gravitational field generated by our source masses by  modulating their position as shown in figure \ref{magiascheme}. A typical elliptical plot 
is shown in figure \ref{ellipses_comparison}, together with the corresponding ellipse of \cite{Sorrentino10} for comparison. 
We move the masses from the close ($C_1$) to the far ($C_2$) configuration and viceversa every $\sim27$ minutes, corresponding to 720 measurement cycles of 1.9\,s each plus a dead time of $\sim5$ minutes to translate the masses. 
We reverse the direction of the $k_e$ vector after each launch, in order to cancel possible $k_{e}$-independent systematic errors, such as those arising from II order Zeeman shift and I order light shift \cite{louchet2011}. We thus obtain two ellipses of 360 points each, corresponding to direct and reverse $k_e$. We fit each set of 360 points to an ellipse, and from each pair  of ellipses we determine the angle $\Phi_n(i)=\Phi_n^{dir}(i)-\Phi_n^{rev}(i)$ as the difference between direct and reverse angles, and the standard error $\delta\Phi_n(i)=\sqrt{{\delta\Phi_n^{dir}(i)}^2+{\delta\Phi_n^{rev}(i)}^2}$. Here $n=1,2$ corresponding to the two configurations of source masses. 
From each couple $\{\Phi_1(i),\Phi_2(i)\}$ a value
for the rotation angle $\Delta\Phi(i)=\Phi_1(i)-\Phi_2(i)$ due to the position of the source masses  can be obtained. 

\begin{figure}
\includegraphics[width=0.5\textwidth]{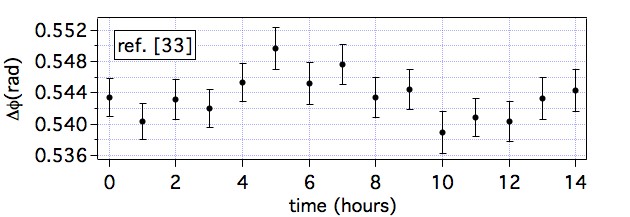}
\includegraphics[width=0.5\textwidth]{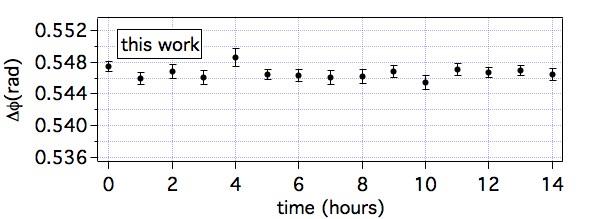}%
\caption{\label{comparison} Differential phase  $\Delta\Phi$ measured over 14 hours; the upper plot corresponds to the experiment status described in \cite{Sorrentino10}; lower plot 
corresponds to the present state of the apparatus.}%
\end{figure}

Figure \ref{comparison} shows two measurements of the differential interferometric phase $\Delta\Phi(i)$ on a period of 14 hours. The upper plot  corresponds to the experimental status described in \cite{Sorrentino10}; the lower plot 
corresponds to the present state of the apparatus. The average values 
  are not comparable, since the positions $C_1$ and $C_2$ of source masses were modified between the two measurements.
The error on each point is $\delta\Delta\Phi(i)\simeq0.74$\,mrad. 
The weighted average of data has a statistical error of 200\,$\mu$rad with a $\chi^2$ of 15. This corresponds to an uncertainty of $3.5\times10^{-4}$ after an integration time of 14\,hours, expecting to reach the $10^{-4}$ level in about one week of continuous measurement.

\section{Conclusions}

We studied the sensitivity and long term stability of 
a gravity gradiometer based on Raman atom interferometry. 
We discussed the influence of the most relevant experimental parameters, in particular for a measurement of the Newtonian gravitational constant.
Our experiment can run continuously for several days,
showing a reproducibility of the gravity gradient measurement at the level of $\sim5\times10^{-9}$\,s$^{-2}$ on the time scale of a few weeks. Our measurement of the differential gravity signal of source masses can reach a statistical uncertainty of  $3.0\times10^{-4}$ 
after $\sim10$ hours of integration time.










%




%











%







\begin{acknowledgments}

This work was supported by INFN (MAGIA experiment) and EU (iSense STREP project Contract No. 250072).
The authors acknowledge M. Depas, M. Giuntini, A. Montori, R. Ballerini, M. Falorsi for technical support.
\end{acknowledgments}


%

\end{document}